\documentclass[11pt]{article}

\usepackage{times}
\usepackage{epsfig}
 
\title{Scaling laws in the functional content of genomes:\\
Fundamental constants of evolution?}
 
\author{Erik van Nimwegen\\
\\
\normalsize{Division of Bioinformatics, Biozentrum, University of Basel,}\\
\normalsize{Klingelbergstrasse 50-70, 4056 Basel, Switzerland}\\
\\
\normalsize{E-mail:  erik.vannimwegen@unibas.ch}
}

\date{}
\begin{document}
\maketitle

\section{Power laws in genomic quantities}

A few years ago it was noticed that the distributions of gene-family
sizes in fully-sequenced genomes follow power-law distributions
\cite{Huynen&vanNimwegen1998,Gerstein97}. Since then different authors
have shown that there is in fact a large array of genomic features
that show power law distributions. Almost all of these concern the
distributions of genomic features within a single genome. For
instance, it is shown in \cite{LuscombeEtAl2002} that the number of
genomic occurrences of DNA words, protein folds, superfamilies, and
families all follow power-law distributions. Power-law distributions
are also found in the structure of the `protein universe'
\cite{Koonin&Wolf&Karev2002}; the number of protein families per fold
is power-law distributed, and so is the number of different assigned
biological functions per fold \cite{LuscombeEtAl2002}. Power-laws also
appear in the structure of cellular interaction and regulatory
networks. For example, the number of genes that a given gene interacts
with is power-law distributed. This holds both when one defines
'interaction' between genes on the level of the proteins that they
encode \cite{UetzEtAl2000,ItoEtAl2001,Maslov&Sneppen2002} or if one
defines it at the level of co-regulation of the expression of the
genes \cite{StuartEtAl2003}. The experimental data on transcription
regulatory networks is rather incomplete but they also suggest that
the number of genes regulated per transcription factor might have
power-law tails \cite{GuelzimEtAl2002,MiloEtAl2002}. Finally,
power-laws also appear in cellular metabolic networks; the number of
substrates that any given substrate interacts with is power-law
distributed \cite{JeongEtAl2000,Wagner&Fell2001}.

\section{Comparing genomic features across genomes}

Note that almost all the power-law distributions just mentioned refer
to statistics that are taken over a {\em single} genome or cellular
network. The statistics of genomic features {\em across} genomes has
been much less (if at all) investigated. To a large extent this may be
because until recently there simply weren't enough fully-sequenced
genomes to obtain meaningful statistics across genomes. However, this
situation is changing rapidly. 

There are currently about $150$
fully-sequenced microbial genomes in genbank and this number appears
to grow exponentially as I have shown in \cite{vanNimwegenTIG2003}. 
\begin{figure}[htbp]
\centerline{\epsfig{file=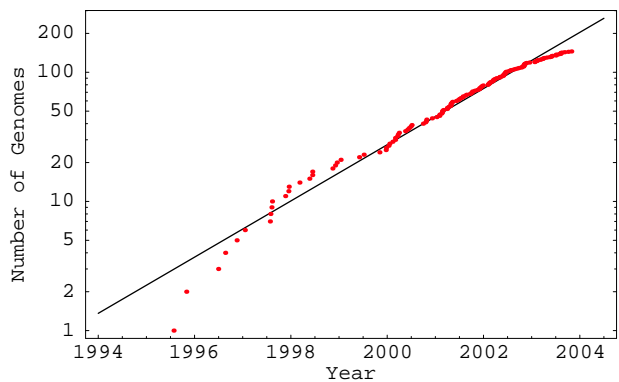,height=5cm}}
\caption{The number of fully-sequenced microbial genomes submitted to
  the genbank database as a function of time (in years). The vertical
  axis is shown on a logarithmic scale. The black line is the
  least-square fit to an exponential. $n = 2^{(t-1993.4)/1.38}$. }
\label{genome_number_scaling}
\end{figure}
Figure \ref{genome_number_scaling} shows an updated plot of the number
of fully-sequenced microbial genomes as a function of time (see
Methods section). The current number of available microbial genomes is
only large enough to allow for meaningful cross-genome comparisons of
the most basic statistics of gene-content and organization and this is
what I will focus on in this chapter. However, the exponential fit in
figure \ref{genome_number_scaling} predicts that the number of
sequenced genomes doubles roughly every $17$ months. This implies that
by $2010$ we may have as many as $3000$ fully-sequenced microbial
genomes available. It is therefore clear that much more detailed
comparative genomic analyses than the ones presented in this chapter
will become possible over the next decade.

In \cite{vanNimwegenTIG2003} I compared the number of genes in
high-level functional categories across all sequenced genomes and
showed that they follow power-laws as a function of the total number
of genes in the genome. In this chapter I will recapitulate these
results and augment them in several ways. In particular, I have
extended the analysis to all functional categories that are
represented by at least $1$ gene in each bacterial genome and have
recalculated the observed exponents based on the latest genomic
data. Second, I will go into more detail regarding the implications of
the observed scaling laws for the general organization of gene-content
across genomes and discuss an evolutionary model that relates the
observed scaling behavior in gene content to fundamental constants of
the evolutionary process. Finally, I will discuss theoretical
explanations for the category of transcription regulatory genes which
shows approximately quadratic scaling with the total number of genes
in the genome.

\section{Scaling in functional gene-content statistics}

To count and compare the number of genes in different functional
categories for all sequenced genomes one needs to first define a set
of functional categories and then annotate all genomes in terms of
these functional categories. I used the biological process hierarchy
of the Gene Ontology \cite{geneontologypaper} to define functional
categories, and Interpro annotations of fully-sequenced genomes to
associate genes with GO categories. The details of the annotation
procedure are described in the Methods section. The result is a count
of the number of genes associated with each of the GO categories in
the biological process hierarchy for each of the sequenced genomes.

The set of genomes in this study consists of $116$ bacteria, $15$
Archaea, and $10$ Eukaryota. In this chapter I will focus solely on
the bacterial data since this is the only kingdom for which there is
sufficient data to obtain meaningful statistics. The reader is
referred to \cite{vanNimwegenTIG2003} for a discussion of the observed
scaling laws in Archaea and Eukaryota. 

\begin{figure}[htbp]
\centerline{\epsfig{file=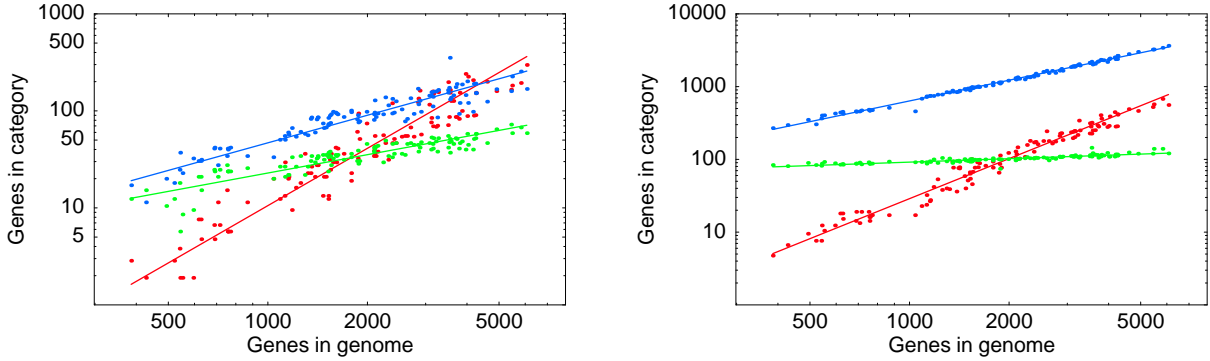,height=5cm}}
\caption{The number of genes involved in (left panel) signal
  transduction (red), carbohydrate metabolism (blue), and DNA repair
  (green), and (right panel) any biological process (blue),
  transcription regulation (red), and protein biosynthesis (green) as
  a function of the total number of genes in the genome that have any
  functional annotation at all. Each colored dot represents the counts
  for a single genome. Both axis are shown on a logarithmic scale. The
  straight lines show power-law fits: (left panel) $n = 0.000015
  g^{1.95}$ (red), $n = 0.063 g^{0.96}$ (blue), $n = 0.37 g^{0.605}$
  (green), and (right panel) $n = 0.000095 g^{1.83}$ (red), $n= 0.92
  g^{0.95}$ (blue), and $n= 30.8 g^{0.16}$ (green).}
\label{example_scaling_fig}
\end{figure}

There are $154$ GO categories in the biological process hierarchy that
have at least $1$ associated gene in each of the $116$ bacterial
genomes. I will refer to these categories as the `ubiquitous'
categories. The results for a selection of $6$ of these ubiquitous GO
categories are shown in figure \ref{example_scaling_fig}. The figure
shows the dependence of the number of genes in each category on the
total number of genes with annotation in the genome.
The left panel shows the categories ``signal transduction'' (red),
``carbohydrate metabolism'' (blue), and ``DNA repair'' (green), while
the right panel shows the categories ``transcription regulation''
(red), ``biological process'' (blue), and ``protein biosynthesis''
(green). Each dot represents the counts in a single bacterial genome
and both axes in Fig. \ref{example_scaling_fig} are shown on a
logarithmic scale. Note that in \cite{vanNimwegenTIG2003} a similar
plot was shown but with the horizontal axis representing the total
number of genes rather than the total number of annotated genes. Since
the total number of annotated genes is to a very good approximation
proportional the total number of genes, the results are virtually
identical whether one uses the total number of genes or the total
number of annotated genes. I decided to use the total number of
annotated genes on the horizontal axis in figure
\ref{example_scaling_fig} partly to illustrate this fact. In addition,
the genome size in bacteria is also to a very good approximation
proportional to the total number of genes in the genome. Thus, if we
had used the genome size instead of the number of annotated genes on
the horizontal axis Fig. \ref{example_scaling_fig} would again have
looked virtually identical.

The dots of each color in Fig. \ref{example_scaling_fig} fall
approximately on a straight line. Thus, the logarithms of the number
of genes $n_c$ in a category $c$ and the total number of genes $g$ (or
the number of annotated genes or the genome size) are approximately
linearly related:
\begin{equation}
\log(n_c ) = a_c \log(g) + b_c.
\end{equation}
In other words, the number of genes $n_c$ in a category increases as a
{\em power-law} in the total number of genes $g$:
\begin{equation}
\label{powerlaw_equation}
n_c = \lambda_c g^{\alpha_c}.
\end{equation}

For the $6$ functional categories shown, the exponents $\alpha_c$ of
the best power-law fits are indicated in the figure caption. The fits
were obtained using the procedure described in the Methods
section. The exponents range from $\alpha = 0.16$ for protein
biosynthesis to $\alpha = 1.95$ for signal transduction.

\begin{figure}[htbp]
\centerline{\epsfig{file=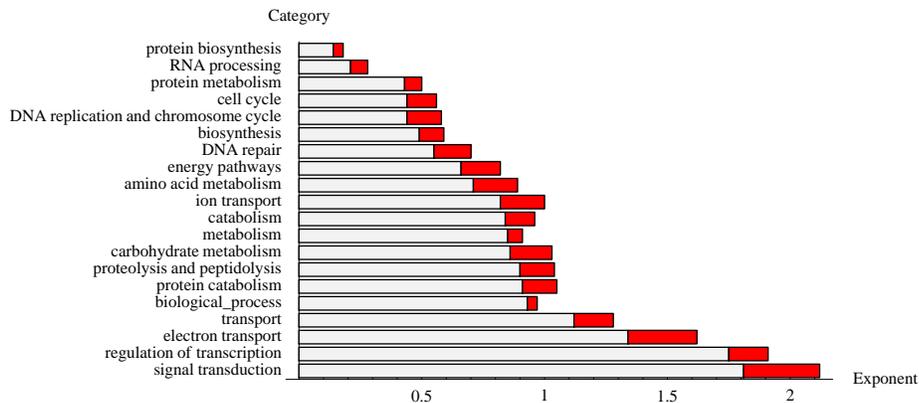,height=8cm}}
\caption{The $99$\% posterior probability intervals for the scaling
  exponents of a selection of functional categories. The functional
  categories are indicated on the left of each bar and the red section
  of the bar indicates the $99$\% posterior probability interval for
  the exponent of that category. The categories are ordered from top
  to bottom by increasing lower-bound of the posterior probability
  interval.}
\label{example_exponents}
\end{figure}

To further show the range and variation of the observed exponents
Fig. \ref{example_exponents} shows the inferred exponents and their
$99$\% posterior probability intervals for a selection of $20$
functional categories.
The exponents range from close to zero to roughly $2$. Note that, for
a category $c$ with exponent $\alpha_c$ the relative {\em proportion}
$p_c$ of genes in the genome scales as $p_c = \lambda_c
g^{\alpha_c-1}$. That is, when $\alpha_c < 1$ the proportion of genes
in the category will decrease with genome size, while for $\alpha_c >
1$ the proportion of genes in the category will increase with genome
size. Thus, for a category $c$ with exponent close to $2$, the
proportion $p_c$ will increase almost linearly with genome size. The
behavior of different categories thus ranges from categories where the
number of genes is almost constant with genome size ($\alpha_c \approx
0$), to categories where the proportion of genes in the genome
increases linearly with genome size ($\alpha_c \approx 0$).

The general picture that emerges from Fig. \ref{example_exponents} is
 that the proportion of genes in essential low-level functional
 categories such as protein biosynthesis and DNA replication decreases
 with genome size, whereas the proportion of genes that play
 regulatory roles such as genes involved in signal transduction and
 transcription regulation increases approximately linearly with genome
 size. In between these extremes is a number of categories, including
 different metabolic functions, for which the exponent is roughly $1$,
 indicating that the genomic percentage of genes in these categories
 is roughly independent of genome size.

\subsection{Upper bound on genome size}

The observed quadratic scaling of the number of regulatory genes with
the total number of genes in the genome obviously cannot extend to
arbitrarily large genome sizes. If we extend the red curves,
corresponding to ``transcription regulation'' and ``signal
transduction'', in figure \ref{example_scaling_fig} to the right, we
will eventually reach a point where the number of signal transducers
and transcription regulators would be larger than the total number of
genes in the genome and this is obviously impossible. Thus, if all
bacterial genomes obey the relations indicated in our figure, there
must be an upper bound on bacterial genome size. A naive upper bound
is obtained by demanding that the number of genes in any category is
less than the total number of genes in the genome, i.e. $n_c =
\lambda_c g^{\alpha_c} \leq g$. If one substitutes the values of the
fits for transcription regulation into this equation one obtains an
upper bound of approximately $g \leq 70000$ genes. A tighter upper
bound is obtained when one demands that $n_c$ cannot increase by more
than $1$ gene when $g$ is increased by $1$ gene, i.e. $\lambda_c
(g+1)^{\alpha_c} \leq \lambda_c g^{\alpha_c} +1$\footnote{Note that in
principle nothing keeps a genome from increasing $n_c$ by more than
$1$ gene when $g$ is increased by $1$ gene. That is, some genes in
other categories than $c$ may be removed and replaced by genes in
category $c$.}. One then obtains an upper bound of $g \leq 34000$ when
the values for transcription regulation are substituted. In
\cite{CroftEtAl2003} an upper bound is derived by assuming that the
number of genes involved in transcription regulation has to be less
than half of the genome size, i.e. $n_c \leq g/2$. With our fit this
leads to an upper bound of about $g \leq 30000$.

It is clear that all these upper bounds substantially overestimate the
approximately $10000$ genes of the largest observed bacterial genomes
and that a more realistic theory is needed to plausibly explain the
apparent size-constrained on bacterial genomes. In this regard it is
also interesting to note that in all the upper bounds just proposed,
the proportion of genes that are transcription regulators is at least
$50$\% at the maximal genome size, whereas the percentage is at most
$11$\% in the currently sequenced bacterial genomes.

\subsection{Consequences for the topology of the transcription regulatory network}

The approximately quadratic scaling of the number of transcription
regulatory genes also has some interesting consequences for the
structure of the transcription regulatory network as a function of
genome size. The class of transcription regulatory genes consists for
the most part of DNA-binding transcription factors that regulate
transcription through the binding to specific regulatory motifs in
intergenic regions. We can imagine the transcription regulatory
network by a set of arrows pointing from each of the regulators to
each of the genes that they regulate. The total number of arrows in
this network for a genome of size $g$ is the product of the number of
genes $g$ times the average number of incoming arrows per gene
$\langle r(g) \rangle$, i.e. $\langle r(g) \rangle$ represents the
average number of different regulators regulating each gene in a
genome of size $g$. Note that we can also write the total number of
arrows as the number of transcription factors $n_{\rm tr}(g)$ in a
genome of size $g$ times the average number of genes $\langle n(g)
\rangle$ that each regulator regulates. We thus have
\begin{equation}
\label{reg_size_scaling}
n_{\rm tr}(g) \langle n(g) \rangle = \langle r(g) \rangle g
\Leftrightarrow \frac{\langle r(g) \rangle}{\langle n(g) \rangle}
\propto g,
\end{equation}
where the equation on the right follows from the quadratic scaling of
the number of regulators: $n_{\rm tr}(g) \propto g^{2}$. To elucidate
what the equation on the right implies, let us consider what follows
if we assume that either $\langle n(g) \rangle = {\rm constant}$ or
$\langle r(g) \rangle = {\rm constant}$. In the first case the average
number of genes regulated per transcription factor, i.e. the regulon
size, is independent of genome size. In that case the average number
of different transcription factors $\langle r(g) \rangle$ regulating
each gene must be increasing linearly with genome size, i.e. $\langle
r(g) \rangle \propto g$ If on the other hand $\langle r(g) \rangle$
were constant with genome size, then the average regulon size $\langle
n(g) \rangle$ should be decreasing with genome size, i.e. $\langle
n(g) \rangle \propto 1/g$. Between these two extremes there is a
continuum of solutions where $\langle r(g) \rangle$ increases more
slowly, and $\langle n(g) \rangle$ decreases more slowly, but such
that still $\langle r(g) \rangle/\langle n(g) \rangle \propto g$.

\begin{figure}[htbp]
\centerline{\epsfig{file=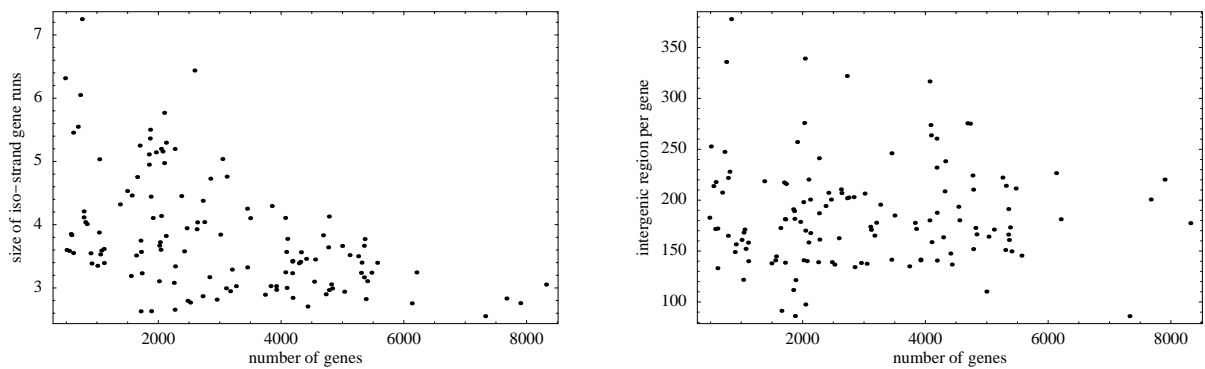,height=5cm}}
\caption{The average length of runs of genes that are transcribed from
  the same strand (left panel) and the average number of intergenic
  bases per gene (right panel) as a function of the total number of
  genes in the genome. Each dot corresponds to a bacterial genome
  in the NCBI database}
\label{operons_intergenic_fig}
\end{figure}

There is currently very little data to decide if real regulatory
networks are closer to the limit where $\langle r(g) \rangle$
increases linearly, or closer to the limit where $\langle n(g) \rangle
\propto 1/g$.  One piece of indirect evidence is the dependence of the
number of operons and the amount of intergenic region on genome
size. If the average number of transcription factors $\langle r(g)
\rangle$ regulating each gene were to increase with genome size, then
one might expect that, as genome size increases, the average operon
size should decrease and that the amount of intergenic region per gene
should increase. It is of course a nontrivial task to identify the
number of operons from genome sequence alone. However, as a proxy we
may consider runs of consecutive genes that are located on the same
strand of the DNA (see \cite{Cherry2003} for a method of estimating
operon number using this statistic). Since all genes in an operon
necessarily have to be transcribed in the same direction, a decrease
in operon number would likely be reflected by a decrease in the
average length of runs of consecutive genes on the same strand. Figure
\ref{operons_intergenic_fig} (left panel) shows this average length of
iso-strand genes as a function of genome size for all bacterial
genomes that are currently in the NCBI database of fully-sequenced
genomes.
The figure suggests a slight decrease of the length of these runs,
consistent with what was reported in \cite{Cherry2003}, but there is a
large amount of variation and the trend is far from convincing,
i.e. $r^2 = 0.23$ under simple regression. Note also that the drop in
operon size is at most a factor of two between the largest and
smallest genomes, while total gene number increases almost a factor of
$20$. The right panel shows the average number of intergenic bases per
gene as a function of the total number of genes in the genome. In this
case a correlation between genome size and the amount of intergenic
region is completely absent, i.e. $r^2 = 0.0005$. Thus these two
statistics provide little evidence that $\langle r(g) \rangle$
increases substantially with genome size. However, one cannot exclude
the possibility that in small genomes a large proportion of genes is
not transcriptionally regulated at all, and that as genome size
increases this proportion drops dramatically. This would still lead to
a substantial increase of $\langle r(g) \rangle$ with genome size. It
does seem plausible, however, that larger genomes may have a larger
number of `specialized' regulons that typically regulate a smaller
number of genes compared to the more general regulators that one
expects to find in all organisms, and that $\langle n(g) \rangle$ thus
decreases with genome size.

\subsection{Quality of the fits}

\begin{figure}[htbp]
\centerline{\epsfig{file=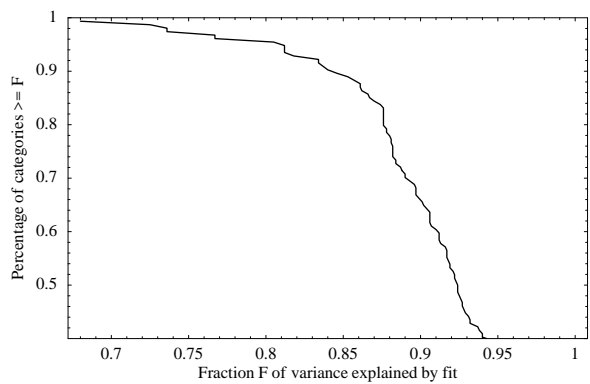,height=5cm}}
\caption{Cumulative distribution of the quality of the power-law fits
  for the $154$ ubiquitous functional categories. The horizontal axis
  shows the fraction $F$ of the variance in the data explained by the
  fit, and the vertical axis shows the percentage of categories that
  have at least a fraction $F$ of their variance explained by the
  fit.}
\label{quality_fits}
\end{figure}

The power-laws observed in figure \ref{example_scaling_fig} are
observed for the large majority of the $154$ ubiquitous functional
categories. I assessed the quality of the power-law fits by a measure
$F$ that measures the fraction of the variance in the data that is
explained by the power-law fit (see Methods). Figure
\ref{quality_fits} shows the cumulative distribution of $F$ for all
$154$ ubiquitous functional categories.
As can be seen from Fig. \ref{quality_fits} about two thirds of the
categories have more than $90$\% of the variance explained by the
fit. More than $95$\% of the categories have more than $80$\% of the
variance explained by the fit. We thus see that most of ubiquitous
functional categories follow scaling laws like the ones shown in Fig.
\ref{example_scaling_fig}.

\section{Principle Component Analysis}

Instead of considering the scaling behavior of one functional category
at a time, we can consider all functional categories at once. We can
consider a $154$-dimensional 'function space' in which each axis
represents an ubiquitous functional category, and represent the
'functional gene content' of a genome by a point in this
$154$-dimensional space. That is, if we number the categories $1$
through $154$, and $n_c$ is the number of genes in category $c$, then
$\vec{n} = (n_1,n_2,\ldots,n_{154})$ represents the functional gene
content of the genome as a point in the function space. The set of all
sequenced genomes thus forms a scatter in this function space.

\begin{figure}[htbp]
\centerline{\epsfig{file=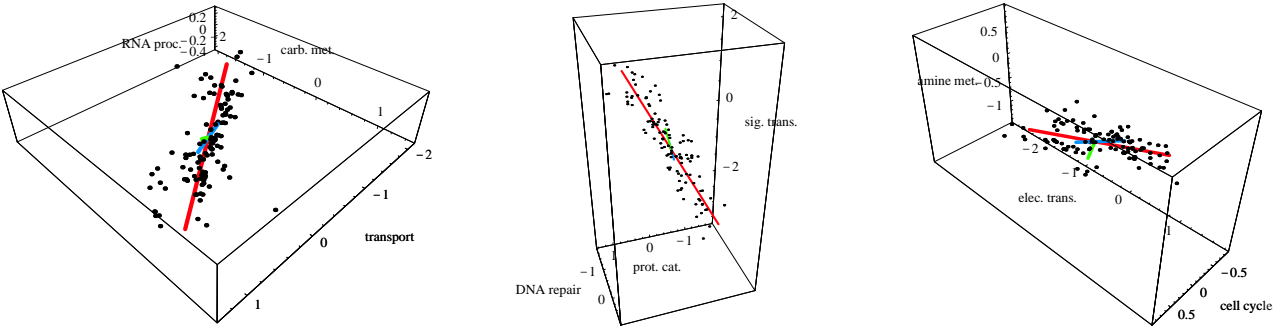,height=4.5cm}}
\caption{Three 'slices' through the scatter of genomes in function
space. Each point corresponds to a genome, and its components along
the three axes correspond to the logarithms of the gene numbers in
each of three categories that the axes represent. The scatter is shown
with respect to the categories 'carbohydrate metabolism', 'transport',
and 'RNA processing' (left panel); 'protein catabolism', 'DNA repair',
and 'signal transduction' (middle panel); and 'cell cycle', 'electron
transport', and 'amine metabolism' (right panel). The first three
principle component axes are shown as the red, blue, and
green lines in the figures (see below).}
\label{three_intersec}
\end{figure}

We can now ask what the shape is of this cloud in function space. To
do this, it is convenient to again consider all axes on logarithmic
scales. That is, we consider the scatter of points $\vec{x}$ with $x_c
= \log(n_c)$. The results in the previous section showed that, to a
good approximation, almost all function categories obey the linear
equations
\begin{equation}
\label{linear_eq}
x_c = \alpha_c \log(g) + \beta_c,
\end{equation}
where $g$ is the total number of genes in the genome. If this equation
were to hold exactly for all categories, then the numbers $x_c$ and
$x_{\tilde{c}}$, for any two categories $c$ and $\tilde{c}$, would
also be linearly related:
\begin{equation}
x_c = \frac{\alpha_c}{\alpha_{\tilde{c}}} x_{\tilde{c}} + \beta_c
-\frac{\alpha_c}{\alpha_{\tilde{c}}} \beta_{\tilde{c}}.
\end{equation}
Thus, the statement that the equation (\ref{linear_eq}) holds for all
categories is equivalent to the statement that the scatter of points
in function space fall on a straight line. Of course, since equation
(\ref{linear_eq}) holds only approximately for each category, the
scatter of points falls only approximately on a line. To
illustrate this figure \ref{three_intersec} shows three projections of
the scatter of points onto three-dimensional subspaces each
representing three functional categories.
That is, each of these three figures shows the scatter of points with
respect to only three of the $154$ axes.  The functional categories
that were used for the projections are indicated in each of the plots
and in the caption. As can be seen, the scatter of points indeed
approximately falls on a straight line in each of these projections.

 The extent to which the scatter of points falls on a line can be
quantified most directly using principle component analysis
\cite{Joliffe1986}. In principle component analysis one aims to
represent a scatter of points in a high-dimensional space by a scatter
in a much lower dimensional space. To this end it finds an ordered set
of orthogonal coordinate axes in the $n$-dimensional space that have
the property that, for any $m \leq n$, the sum of the squared
distances of the data points to their projections on the first $m$
axes of the coordinate system is minimized (i.e. no set of $m$ axes
has a lower sum of squared distances). That is, the first principle
axis is chosen such that minimizes the average square-distance of the
data points this axis is minimized. The second principle axis is
chosen such that the average square-distance of the data points to the
surface spanned by the first and second principle axis is
minimized. And so on for the further principle components.

The ability of this coordinate system to represent the scatter of
points is again measured by the fraction of the variance in the data
that is captured by consecutive axes. For the scatter of $116$
bacterial genomes in the function space of $154$ ubiquitous
categories, the first principle axis captures $22$\% of all the
variance in the data. The second principle axis captures $6$\% of the
variance, the third $5$\% of the variance, etcetera. Thus, the amount
of variance explained drops sharply between the first and second
principle axis. After that, the amount of variance explained by the
data decreases smoothly, dropping between $5$ and $10$ percent between
consecutive axes. As many as $55$ axes are needed to cover $95$\% of
the variance in the data. These statistics suggest that only the first
principle axis captures an essential characteristic of gene-content in
bacterial genomes. Once this largest component of the variance is
taken into account, one needs almost as many dimensions as there are
genomes to explain the remaining variance in gene-content.

\begin{table}
\label{first_three_comp} 
\begin{center}    
\begin{tabular}{|c|c|c|c|} 
\hline   
\multicolumn{4}{|c|}{\bf First principle axis}\\
\hline   
\hline
top and bottom categories. & comp. & top and bottom genomes & comp. \\
\hline
cell communication & 1.91 & Streptomyces coelicolor & 0.15\\
signal transduction & 1.91 & Streptomyces avermitilis & 0.14\\
regulation of transcription & 1.88 & Bradyrhizobium japonicum & 0.14 \\
electron transport & 1.46 & Rhizobium loti & 0.14\\
 & & & \\
hydrogen transport & 0.25 & Buchnera aphidicola & -0.19\\
protein biosynthesis & 0.17 & Mycoplasma pneumoniae & -0.22 \\
ATP metabolism & 0.10 & Ureaplasma parvum & -0.24\\
rRNA modification & 0.06 & Mycoplasma genitalium & -0.24\\
\hline
\hline
\multicolumn{4}{|c|}{\bf Second principle axis}\\
\hline
porphyrin metabolism & 0.28 & Mycobacterium leprae & 0.16\\
fatty acid metabolism & 0.20 & Prochlorococcus marinus & 0.16\\
carboxylic acid biosynthesis & 0.19 & Wigglesworthia glossinidia
brevipalpis & 0.16 \\
heterocycle metabolism & 0.14 & Candidatus Blochmannia floridanus &
0.15\\
 & & & \\
DNA alkylation & -0.16 & Mycoplasma penetrans & -0.22\\
epigenetic regulation of gene expression & -0.17 & Borrelia burgdorferi & -0.22\\
nucleoside metabolism & -0.17 & Ureaplasma parvum & -0.22\\
aspartyl-tRNA aminoacylation & -0.17 & Mycoplasma pulmonis  & -0.3\\
\hline
\hline
\multicolumn{4}{|c|}{\bf Third principle axis}\\
\hline
protein secretion & 0.31 & Borrelia burgdorferi & 0.46\\
cell communication & 0.15 & Chlamydia trachomatis & 0.26\\
signal transduction & 0.13 & Treponema pallidum & 0.23\\
DNA dependent DNA replication & 0.09 & Chlamydia muridarum & 0.23\\ 
 & & & \\
amino acid biosynthesis & -0.16 & Streptococcus pneumoniae & -0.13\\
ribonucleotide metabolism & -0.17 & Prochlorococcus marinus & 0.14\\
purine nucleotide metabolism & -0.18 & Wigglesworthia glossinidia
brevipalpis & -0.15\\
ATP metabolism & -0.18 & Candidatus Blochmannia floridanus & -0.19\\
\hline
\end{tabular}
\caption{Summary of the first three principle axes. Each principle
axis is a vector $\vec{a}$ in function space with component $a_c$ the
component of the vector in direction of category $c$. For each
principle axis the $4$ categories with the highest, and $4$ categories
with the lowest $a_c$ are shown (omitting redundant categories) in the
leftmost column. The second column shows the $a_c$. The location of
each genome in function space can be expressed as a linear combination
of principle axes. For each principle axes shown above the $4$ genomes
with the highest and $4$ genomes with the lowest components along the
axis are shown in column $3$, and the values of the components of
these genomes are shown in column $4$.}
\end{center}
\end{table}
Table \ref{first_three_comp} summarizes the statistics of the first
three principle axes. Each of the axes is a vector $\vec{a}$ whose
direction in function space is reflected by the relative sizes of the
components $a_c$. In the leftmost column of table
\ref{first_three_comp} I have listed, for each axis, the $4$
categories with the highest components $a_c$ and the $4$ categories
with the lowest components $a_c$ (redundant categories were
omitted). The values of these components $a_c$ are shown in the second
column. The projection of all genomes onto one of the principle axes
gives another vector $\vec{v}$ where $v_g$ is the component of genome
$g$ in the direction of the principle axis. The third column in table
\ref{first_three_comp} shows, for each principle axis, the $4$ genomes
with the highest components $v_g$ and the $4$ genomes with the lowest
components $v_g$. The components themselves are shown in column
$4$. These first three principle axes are also indicated as the red,
blue, and green lines in figure \ref{three_intersec}.

As can be easily derived from equation (\ref{linear_eq}), the
components $a_c$ of the first principle axis correspond precisely to
the inferred exponents of figure \ref{example_exponents}. Thus, the
entire collection of scaling laws is summarized in this single vector
$\vec{a}$. In summary, a large fraction of the variation in functional
gene-content among all bacterial genomes can be summarized into a
single vector which encodes how the numbers of genes in different
functional categories increases and decreases as the total size of the
genome varies. That is, as the genome size increases or decreases the
numbers of genes in each functional category $c$ increase or decrease
as $g^{a_c}$. The vector $\vec{a}$ thus reflects a basic {\em
functional architecture} of gene-content that holds across all
bacterial genomes.

The meaning of the second and third principle axis is less clear, and
given that they only capture a relatively small amount of the variance
it is not clear that they very meaningful at all. For both these axes
the genomes at the extremes tend to be small parasitic
organisms. This suggests that these axes may reflect the different
types of parasitic lifestyles.

\section{Evolutionary interpretation}

What is the origin of the scaling laws discussed in the previous
sections? In this section I will show that the observed scaling laws
in fact suggest that there are fundamental constants in the
evolutionary dynamics of genomes.

Consider a particular genome with numbers of genes $n_c$ in different
functional categories $c$. Consider next the evolutionary history of
this genome. With this I mean that the current genome can be followed
back in time through the life of the cell the genome was taken from,
through the cell division that produced it, through the life of its
ancestral cell, and its ancestors, and so on. In this way the history
of the genome can be traced back all the way to an ancestral
prokaryotic cell from which all currently existing bacteria
stem. During this evolutionary history the numbers $n_c$ have of
coursed increased and decreased in ways that are unknown to us. That
is, there is (unknown) functions of time $n_c(t)$ that describe the
evolution of the numbers of genes in each category $c$ in the genome
under study. Similarly, there will be a function $g(t)$ that describes
the evolutionary history of the total number of genes in the genome.

One can now ask what the constraints the functions $g(t)$ and $n_c(t)$
should obey such that the observed scaling laws hold. To this end it
is convenient to write the dynamics of $n_c(t)$ and $g(t)$ in terms of
effective duplication and deletion rates. That is, we write
\begin{equation}
\frac{d n_c(t)}{d t} = \beta_c(t) n_c(t) - \delta_c(t) n_c(t) \equiv \rho_c(t) n_c(t),
\end{equation}
and
\begin{equation}
\frac{d g(t)}{d t} = \beta(t) g(t) - \delta(t) g(t) \equiv \rho(t) g(t).
\end{equation}
In these equations, $\beta_c(t)$ is the (time-dependent) average
duplication rates of genes in category $c$, $\beta(t)$ is the average
duplication rate of all genes, $\delta_c(t)$ is the average deletion
rate of genes in category $c$, and $\delta(t)$ is the average deletion
rate of all genes. I have also defined the differences of duplication
rates and deletion rates as $\rho_c(t)$ and $\rho(t)$. Notice that,
since in the above equations $\rho_c(t)$ and $\rho(t)$ can be
arbitrary functions of time, any time-dependent function can still be
obtained as the solution of the above equations. Formally solving
these equations we obtain
\begin{equation}
n_c(t) = n_c(0) \exp\left(\int_0^t \rho_c(\tau) d \tau \right) =
n_c(0) \exp\left(\langle \rho_c \rangle t\right),
\end{equation}
and
\begin{equation}
g(t) = g(0) \exp\left(\int_0^t \rho(\tau) d \tau \right) = g(0)
\exp\left( \langle \rho \rangle t\right),
\end{equation}
where I have written the integrals as the {\em time averages} of $\rho_c$
and $\rho$ times the total time $t$. Note that these averages are a
function of the evolutionary history of the genome under study. If we
now express $n_c(t)$ in terms of $g(t)$ we obtain 
\begin{equation}
\label{formal_eq_nc}
n_c(t) = \frac{n_c(0)}{g(0)^{\langle \rho_c \rangle/\langle \rho \rangle}}
\left[g(t)\right]^{\langle \rho_c \rangle/\langle \rho \rangle}.
\end{equation}
Note that, although this equation may appear to already imply the
general scaling relations that were found, in fact it is completely
general and holds for {\em any} set of evolutionary histories $n_c(t)$
and $g(t)$. The equation is nothing more than a way of rewriting the
relation between $n_c(t)$ and $g(t)$ in terms of the averages $\langle
\rho_c \rangle$ and $\langle \rho \rangle$ of this genome's history.

\begin{figure}[htbp]
\centerline{\epsfig{file=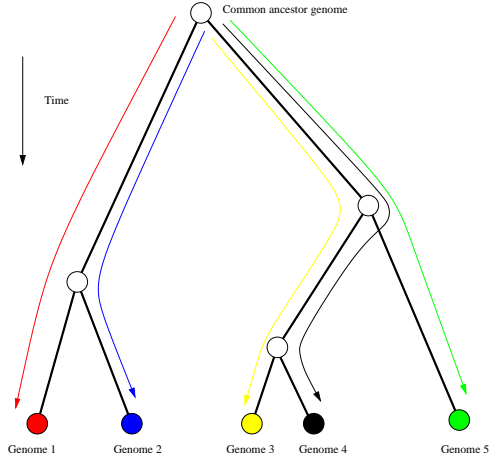,height=6cm}}
\caption{Example of the evolutionary histories of $5$ genomes. Each
  leaf (colored ball) represents a genome. Each white ball corresponds
  to a common ancestral genome of two or more of the
  genomes. The root of the tree is the common ancestor genome of all
  $5$ genomes. Each colored arrow represents the evolutionary history of
  a genome.}
\label{phylogeny_fig}
\end{figure}

The observed scaling relations only follow from (\ref{formal_eq_nc})
if the {\em same} equation holds for all genomes. That is, if the
variables $n_c(0)$, $g(0)$ and $\langle \rho_c \rangle/\langle \rho
\rangle$ are the same for all evolutionary histories. This requirement
is illustrated in figure \ref{phylogeny_fig}.
The figure shows the evolutionary histories of a set of genomes
in a phylogenetic tree. Each leaf of the tree represents one of the
bacterial genomes, and at the root is the common ancestor genome of
all the genomes at the leafs. We thus have a separate equation
(\ref{formal_eq_nc}) for each of the genomes at the leafs. The initial
numbers $n_c(0)$ are $g(0)$ in each of these equations are just the
numbers of genes in category $c$ and in the whole genome of the common
ancestor, and it is therefore clear that the variables $n_c(0)$ and
$g(0)$ are indeed the same for all the equations\footnote{One caveat
is that we implicitly assume that $n_c(t) > 0$ for all $t$. A slightly
more complex treatment is needed for categories that were not present
in the ancestor, or that disappeared and reappeared during the
evolutionary history of some genomes.}.

Much more interesting is the requirement that the ratios $\langle
\rho_c \rangle/\langle \rho \rangle$ are the same for all evolutionary
histories. These different evolutionary histories are indicated as
colored lines in figure \ref{phylogeny_fig}. The requirement that
$\langle \rho_c \rangle/\langle \rho \rangle$ be equal for all
evolutionary histories thus demands that if one takes the average of
the duplication minus deletion rate of genes in category $c$ over the
entire evolutionary history of a genome, and divides this by the
average duplication minus deletion rate of all genes in the genome,
that this ratio always comes out the same, independent of what
evolutionary history is averaged over. That is, the ratios $\langle
\rho_c \rangle/\langle \rho \rangle$ are universal {\em evolutionary
constants} and correspond precisely to the exponents $\alpha_c =
\langle \rho_c \rangle /\langle \rho \rangle$ of the observed scaling
laws.

So what determines $\langle \rho_c \rangle$ and $\langle \rho
\rangle$? It seems reasonable to assume that the rate at which genes
are duplicated and deleted is mostly the result of selection. That is,
as a first approximation we may assume that the rate at which
duplications and deletions are introduced during evolution are
approximately the same for all genes but that different genes have
different probabilities of being selectively beneficial when
duplicated or deleted.  The rate $\rho_c(t)$ is then given by some
overall rate $\gamma$ at which duplications and deletions are
introduced\footnote{In reality the rates at which duplications are
introduced is unlikely to equal the rate at which deletions are
introduced. We ignore this complication for notational simplicity. The
theoretical development with this complication would be analogous.}
times the difference between the fraction of genes $f^{+}_c(t)$ in the
category that would benefit the organism when duplicated and the
fraction of genes $f^-_c(t)$ in the category that would benefit the
organism when deleted:
\begin{equation}
\rho_c(t) = \gamma (f^{+}_c(t)-f^{-}_c(t)).
\end{equation}
We then have
\begin{equation}
\label{alpha_from_rel_rates}
\alpha_c = \frac{\langle \rho_c(t)\rangle}{\langle \rho(t) \rangle} =
  \frac{\langle f^+_c - f^-_c \rangle}{\langle f^+ - f^- \rangle}.
\end{equation}
The fractions $f^+_c$ and $f^-_c$ of course depend on the selective
pressures of the environment. Thus, as one follows the genome through
its evolutionary history, the demand for genes of a certain function
fluctuates up and down, and this will be reflected in fluctuations of
$(f^+_c - f^-_c)$. It seems intuitive clear that the size of the
fluctuations is going to depend strongly on the functional class. That
is, one would expect that demand for genes that provide an essential
and basic function fluctuates relatively little. For instance, one
would expect that even as the selective environment changes it is rare
that existing protein biosynthesis genes become dispensable or that
duplicates of these genes become desirable. On the other hand, the
desirability of transcription regulatory genes and signal transducers
is going to depend crucially on the selective environment in which the
organism finds itself. It is thus not implausible that, if one
averages over sufficiently long evolutionary times that the ratio
$\langle f^+_c - f^-_c \rangle/\langle f^+ - f^- \rangle$ always
reaches the same limit, and that this limit is large for highly
environment-dependent categories such as transcription regulation, and
small for environment-insensitive categories such as protein
biosynthesis and replication. The main open question that remains is
the origin of the precise numerical values of the ratios $\langle
f^+_c - f^-_c \rangle/\langle f^+ - f^- \rangle$ for different
functional categories.

\subsection{The exponent for transcription regulatory genes}

The exponent $\alpha_c$ for the category of genes involved in
transcription regulation is close to $2$, with the $99$\% posterior
probability interval running from $1.72$ to $1.95$. Thus, even though
the current data suggests that the exponent is slightly less than $2$,
it is tempting to think that the scaling might be simply
quadratic. This is especially tempting given that this allows one to
speculate more easily about the origin of this exponent. That is, it
is easier to theorize about a quadratic scaling law then about a
scaling law with exponent $1.83$. Some theoretical explanations for
the quadratic scaling of transcription regulators have recently been
put forward \cite{CroftEtAl2003}. In this section I want to discuss
these proposals and contrast it with my own suggestions in this regard
in \cite{vanNimwegenTIG2003}.

One of the first things that of course come to mind when attempting
to explain a scaling of the form $n_c \propto g^2$ is that in a genome
with $g$ genes, the number of {\em pairs} of genes scales precisely as
$g^2$. That is, there are on the order of $g(g-1)/2$ possible interactions
in a genome with $g$ genes. Therefore, if one can find any way to
argue that the number of transcription regulators should be
proportional to the number of pairs of genes, this would provide an
explanation. In a recent paper \cite{CroftEtAl2003} Croft and
coworkers put forward two models for the observed (approximately)
quadratic scaling of the number of transcription regulatory genes that
both in essence attempt to argue that the number of regulators should
be proportional to the number of potential interactions between genes,
i.e. the number of pairs of genes in the genome. 

Although it is attractive to seek such a simple combinatorial
explanation I believe that a simple survey of the known regulatory
role real transcription factors play shows that such models are in
fact highly implausible. If each regulator were to somehow
`correspond' to one or more pairwise interactions of genes then one
would expect that the role of most regulatory genes would be to ensure
that certain pairs of genes are expressed together or to ensure that
certain pairs are not expressed together.

This is, however, not what is observed. To name just a few known
E. coli regulons: the factor crp responds to the concentration of cAMP
in the cell, and its role is to make the activation of many catabolic
pathways conditional on the absence of cAMP. The factor lexA activates
a number of genes that are involved in the repair of DNA damage in
response to single stranded DNA. PurR represses a set of genes
involved in {\em de novo} purine biosynthesis and responds to the
concentrations of hypoxanthine and guanine. FadR senses the presence
of long chain fatty acyl-coA compounds and in response regulates genes
that transport and synthesize fatty acids. Finally, tyrR regulates
genes that synthesize and transport aromatic amino-acids in response
to levels of phenylalanine, tyrosine, and tryptophan.

In all these cases, the role of the regulator is to sense a particular
signal and to respond to this signal by activating or repressing a set
of genes that implement a specific biological function which is
related to the signal. In none of these example does it appear that
the role of the regulator is in any way related to regulating specific
interactions between pairs of genes. It thus seems that the number of
different transcription regulators that the cell needs is much more a
reflection of the number of different cellular responses to different
environments that the cell is capable of.

In \cite{vanNimwegenTIG2003} I provided a qualitative argument for the
approximately quadratic scaling of the number of transcription
regulatory genes. According to equation (\ref{alpha_from_rel_rates})
the average difference $\langle f^+_c - f^-_c \rangle$ between the
fractions of transcription regulatory genes that would increase
fitness when duplicated and those that would increase fitness when
deleted is almost twice as large as the average difference $\langle
f^+ - f^- \rangle$ between the fractions of all genes that would
increase fitness when duplicated and those that would increase fitness
when deleted. Let us separate the evolutionary history into periods in
which the genome is growing ($f^+ > f^-$) and periods in which it is
shrinking ($f^+< f^-$), and let us for simplicity assume that $f^-_c =
f^- = 0$ during the former periods and $f^+_c = f^+ = 0$ during the
latter periods. We then have that during periods of genome growth
$f^+_c \approx 2 f^+$ and that during periods of genome shrinkage
$f^-_c \approx 2 f^-$. That is, during genome growth transcription
regulators are twice as likely to lead to fitness increase when
duplicated as average genes, and during genome shrinkage transcription
regulators are twice as likely to lead to fitness increase when
deleted.

I want to suggest that the origin of this factor $2$ is the
switch-like function of transcription factors. Imagine a gene that has
just emerged through a duplication. Originally the duplicate will be
the same as its parent. At that point, the main change caused by this
duplication that may affect fitness is a change in the dosage of the
gene, i.e. from one to two copies. One would expect that the
probability for this dosage change to have strong deleterious effects
is approximately equal for transcription regulators as for genes in
general. If the duplicated gene is to get fixed in the genome on a
longer time scale, then a process of mutation and selection should
modify the duplicated gene into a gene that increases the fitness of
the organism. I suggest that the probability for this process to be
successful is twice as high in transcription regulators as in general
genes. In general genes the only avenue for beneficial change is a
change in the molecular function of the gene, e.g. a metabolic gene
may evolve to catalyze a new chemical reaction. Transcription
regulators, however, may evolve to respond to a signal which the cell
was previously insensitive to. They may evolve to affect the state of
the cell both when this signal is present and when the signal is
absent. Since this gives twice as many opportunities for a beneficial
change, one may expect that there is twice as much probability of
success. 

A similar argument hold for the rate of deletion. When deleting a
non-regulatory gene, the cell simply changes to a state without the
gene. When a transcription regulator is removed, one effectively
removes a 'switch' from the genome: the cell becomes insensitive to
the signal that the regulator responded to. But there are again two
'ways' of implementing this sensitivity. The genes regulated in
response to the signal may be turned constitutively on, or
constitutively off. Thus there are two independent ways of removing a
transcription factor, and one would thus expect the effective rate of
transcription factor deletion to be twice the rate of deletion of
general genes. 

These arguments are of course highly speculative and may well turn out
to be incorrect. However, they at least seem consistent with what we
know about transcription regulation in bacteria and the process of
evolution through gene duplication and deletion. Note also that very
similar arguments as the ones just presented for transcription
regulatory genes can be put forward for the category of signal
transducing genes. These are indeed also observed to scale
approximately quadratically with genome size.

I suspect that it will be impossible to come to any solid conclusions
regarding the cause of the approximately quadratic scaling for
transcription regulators and signal transducers until we have better
data on the genome-wide topology of the transcription regulatory and
signal transduction networks in bacteria of differing sizes.

Finally, I note that most of the functional categories have exponents
that do not appear to equal small integers or even simple
rationals. It is thus clear that simple arguments such as the ones
discussed above will not be capable of explaining these
exponents. What determines the average $\langle
f^+_c-f^-_c\rangle/\langle f^+-f^- \rangle$ for these categories is a
fascinating question that at this point is completely open.

\section{Methods}

\subsection{The number of genomes as a function of time}

The data for figure \ref{genome_number_scaling} were obtained by
extracting the submission dates of all the microbial genomes in
genbank at 'ftp://ftp.ncbi.nih.gov/genbank/genomes/Bacteria' from the
'gbs' file for each genome. When there were multiple submission dates
the earliest date was taken. The logarithm of the number of genomes as
a function of time was then fitted to a straight line using standard
least-square regression. This least-square fit gives the number of
genomes $n(t)$ as a function of time (in years) as $n(t) =
2^{(t-1993.4)/1.38}$. The exponential provides a reasonable fit,
i.e. $r^2 = 0.98$ even though the data clearly suggest a decrease in
the rate at which new genomes appear for the last $1$ to $1.5$ years.

\subsection{Gathering functional gene-content statistics}

The numbers of genes in different functional categories for each
sequenced genome were obtained in the following way. Interpro
annotations \cite{interpro_ref} of fully-sequenced genomes were
obtained from the European Bioinformatics Institute
\cite{proteome_database}. Functional categories were taken from the
gene ontology biological process hierarchy \cite{geneontologypaper}. A
mapping from Interpro to GO-categories was also obtained from the gene
ontology website. Using this mapping I gathered, for each GO-category,
all Interpro entries that map to it or to one of its descendants in
the biological process hierarchy. I then counted, for each gene in
each fully-sequenced genome and each GO-category, the number of
Interpro hits $h$ that the gene has to Interpro categories associated
with the GO-category. A gene with many independent Interpro hits that
are associated with the same GO-category is of course more likely to
be a member of the GO-category than a gene with only a single hit. To
quantify this, I chose the probability for a gene to be a member of a
GO-category to which it has $h$ hits to be $1-\exp(-\lambda h)$, with
$h=3$. The results presented in this chapter are largely insensitive
to changes in $h$ (see the discussion in \cite{vanNimwegenTIG2003}).

 In this way the number of genes associated with each GO-category in
each genome were counted. I then selected all GO categories that have
a nonzero number of counts in all bacterial genomes. There are $154$
such ubiquitous GO-categories. I also counted the total number of
genes that have any annotation at all for each genome. These are
defined as genes that have at least one Interpro hit. Further
discussion of this annotation procedure and its robustness can be found
in \cite{vanNimwegenTIG2003}.

\subsection{Power-law Fitting}

In order to fit the data to power-law distributions I used a Bayesian
procedure that is described in \cite{vanNimwegenTIG2003}. The main
advantage of this fitting procedure with respect to simple regression
is that the results are explicitly rotationally invariant. That is,
the best line that the fitting procedure produces doesn't depend on
the orientation of the coordinate axes with respect to the scatter of
data points.

The result is that the posterior distribution $P(\alpha|D)d\alpha$ for the
slope $\alpha$ of the line given the data $D$ is given by
\begin{equation}
\label{form_exponent_posterior}
P(\alpha|D)d\alpha = C \frac{(\alpha^2+1)^{(n-3)/2}d\alpha}{(\alpha^2 s_{xx}-2
\alpha s_{yx} + s_{yy})^{(n-1)/2}},
\end{equation}
where $n$ is the number of genomes in the data, $s_{xx}$ is the
variance in $x$-values (logarithms of the total gene numbers),
$s_{yy}$ the variance in $y$-values (logarithms of the number of genes
in the category), $s_{yx}$ is the co-variance, and $C$ is a
normalizing constant. For each of the $154$ ubiquitous categories we
then calculated the $99$\% posterior probability interval for the
slope from equation (\ref{form_exponent_posterior}).

\subsection{Quality of the power-law fits}

To calculate the quality of the power-law fits, I first log-transform
the data points $(n_i,g_i)$ to $(x_i,y_i) =
(\log(n_i),\log(g_i))$. The best power-law fit is a straight line $y =
\alpha x + \beta$ in the $(x,y)$ plane. For each data point
$(x_i,y_i)$ I then find the distance $d_i(l)$ to this line, and the
distance $d_i(c)$ to the center of the scatter of points. That is,
with $\langle x \rangle$ the average of the $x$-values, and $\langle y
\rangle$ the average of the $y$-values, the distance $d_i(c)$ is given
by
\begin{equation}
d_i(c) = \sqrt{(x_i-\langle x \rangle)^2+(y_i-\langle y \rangle)^2},
\end{equation}
and the distance to the line is given by
\begin{equation}
d_i(l) = \sqrt{\frac{(y_i-\alpha x_i-\beta)^2}{\alpha^2+1}}.
\end{equation}
I then define the quality of the fit $F$ as the fraction of the
variance that is explained by the fit.
\begin{equation}
F = 1- \frac{\sum_i [d_i(l)]^2}{\sum_i [d_i(c)]^2}.
\end{equation}

\bibliography{/home/nimwegen/Scaling/PaperKooninBook/epev.bib}
\bibliographystyle{plain}

\end{document}